\documentclass[submission, Phys]{SciPost}

\begin{document}

\begin{center}{\Large \textbf{
Revealing hidden correlations from complex spatial distributions: Adjacent Correlation Analysis
}}\end{center}

\begin{center}
Guang-Xing Li\textsuperscript{1},

\end{center}

\begin{center}
{\bf 1} South-Western Institute of Astronomical Research, Kunming, Yunnan,
China, 650500

* ligx.ngc7293@gmail.com
\end{center}

\begin{center}
\today
\end{center}


\section*{Abstract}
{
Physics has been transforming our view of nature for centuries. While combining physical knowledge with computational approaches has enabled detailed modeling of physical systems' evolution, understanding the emergence of patterns and structures remains limited. Correlations between quantities are the most reliable approach to describe relationships between different variables. However, for complex patterns, directly searching for correlations is often impractical, as complexity and spatial inhomogeneity can obscure correlations. We discovered that the key is to search for correlations in local regions and developed a new method, adjacent correlation analysis, to extract such correlations and represent them in phase space. When multiple observations are available, a useful way to study a system is to analyze distributions in phase space using the Probability Density Function (PDF). Adjacent correlation analysis evaluates vectors representing local correlations, which can be overlaid on the PDF plot to form the adjacent correlation plot.
These correlation vectors often exhibit remarkably regular patterns and may lead to the discovery of new laws. The vectors we derive are equivalent to the vector field in dynamical systems on the attracting manifold. By efficiently representing spatial patterns as correlation vectors in phase space, our approach opens avenues for classification, prediction, parameter fitting, and forecasting.
}

\vspace{10pt}
\noindent\rule{\textwidth}{1pt}
\tableofcontents\thispagestyle{fancy}
\noindent\rule{\textwidth}{1pt}
\vspace{10pt}

\section{Introduction}{}

Physics is one of the most fundamental scientific disciplines, which describes matter, its fundamental constituents, its motion, and its behavior through space and time. With the development of modern computers, the capability to simulate the evolution of physical systems has been significantly improved, where one often observes the emergence of complex behaviors from systems controlled by sometimes surprisingly simple equations. 
 Examples include the Lorentz system \cite{lorenz1963deterministic}, turbulence \cite{frisch1995tlk}, and Turing equations \cite{1952RSPTB.237...37T}. Towards these complex patterns, a reliable and scientific approach to describe them is a crucial first step to establishing the causal connection between equations, behaviors, and other phenomena in Nature.

Correlation is a widely used and dependable way to describe the relationship between
different quantities. It is both a first step to discovery regularities and
often the final aim of scientific exploration. However, finding
correlations from complex structures is often difficult, as a combination of the
complexity of the structure, lack of a priori knowledge, and spatial inhomogeneity can significantly
undermine the correlations. Under those non-ideal cases, the key to regularity discovery often involves a series of trials and errors, where techniques such as filtering, de-trending, and decompositions are often applied to reveal regularities. 

One overlooked yet effective approach to regularities discovery is to study the correlations between
quantities measured in small, local patches. These strong local correlations appear to be common across datasets. 
 To make this discovery useful, we propose a new method, the
\emph{adjacent correlation analysis}, to extract such locally contained correlations to visualize them in the parameter space. The method retains the capability of the parameter plots in visualizing the global distribution while providing a new way to represent these locally induced correlations that relate to the spatial distribution. These
\emph{adjacent correlation plots} often contain large, coherent subregions
characterized by coherent correlations, indicating similar physics. The regularities of the parameter space revealed by the \emph{adjacent correlation analysis} are both a crucial step in building theory and a starting point for interactive data analysis.

The method bridges a gap between the simplicity of the physics encoded in the governing structures and the complexity of the emergent behaviors. It takes advantage of the fact that a complex system may exhibit simple, predictable behaviors under certain conditions that have been fundamental to previous approaches to modeling and theory-building. This has been discussed under the concept of sloppiness \cite{transtrum2015perspective}, and already used in the \emph{equation-free} approach to modeling \cite{2009ARPC...60..321K}, and other computational approaches such as  \emph{intrinsic low-dimensional manifold}
 \cite{maas1992simplifying} in chemical kinetics and \emph{spectral submanifold}  \cite{Jain_2021} in the study of dynamical systems. The author believes the regularity we search for shares a common origin with the regularity discovered/used in these previous approaches. Compared to those approaches, our method is more orientated towards data visualization and explorations, backed by these deep understandings. The simplicity of our appearance guarantees its wide adoption, and the fact that one can reveal regularities using local correlations should inspire new computational approaches to regularity discovery.


\begin{figure*}[htbp]
    \centering
    \includegraphics[width=1\textwidth]{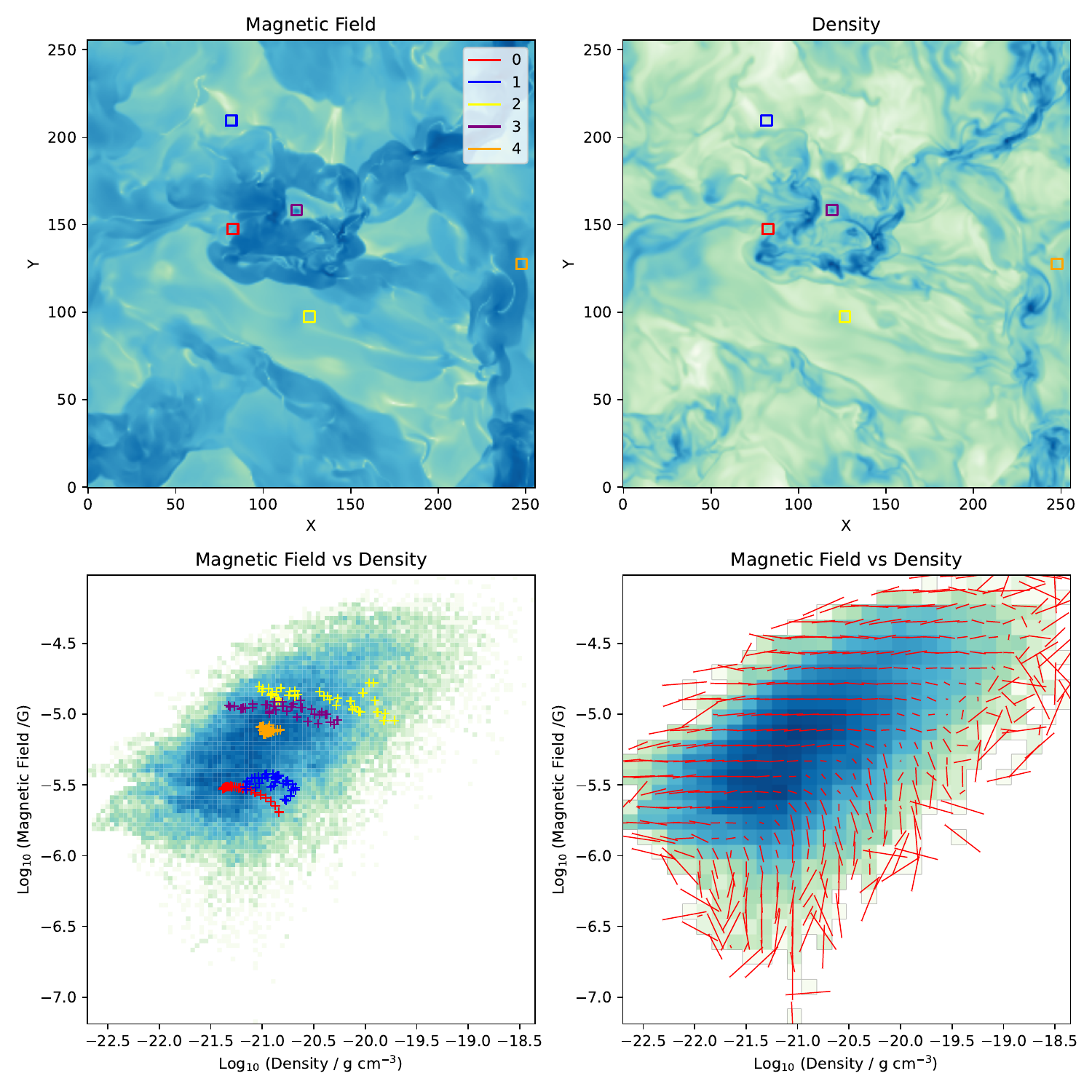}
    \caption{{\bf The phenomenon of Adjacency-Induced Correlations from magnetized turbulence}. In the upper panels, we plot the density $\rho$ and magnetic field $B$ from the simulation, where several small regions are randomly selected. In the lower panels, we plot the magnetic field strength against the gas density. In the lower-left panel, we plot the values of $B$ and $\rho$ from the boxes using symbols of different colors. In the lower-right panel, we use red bars to denote the vectors derived using our method. \label{fig:correlation} } 

\end{figure*} 

 

\begin{figure*}
    \centering
    \includegraphics[width=1\textwidth]{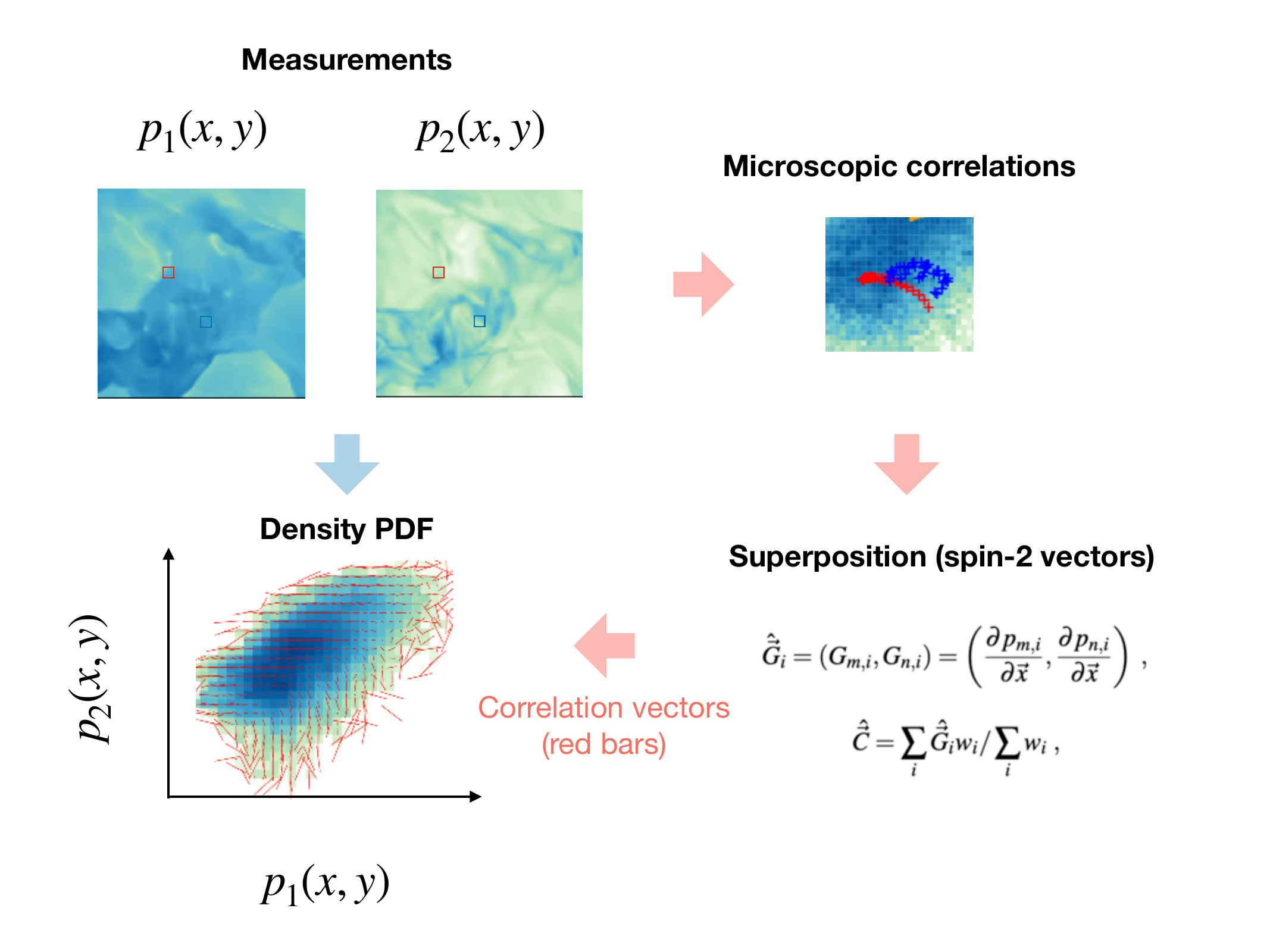}
    \caption{{\bf A diagram illustrating the evaluation of the adjacent correlation vectors. }  \label{fig:illus} } 
\end{figure*}

\begin{figure*}
    \includegraphics[width=1\textwidth ]{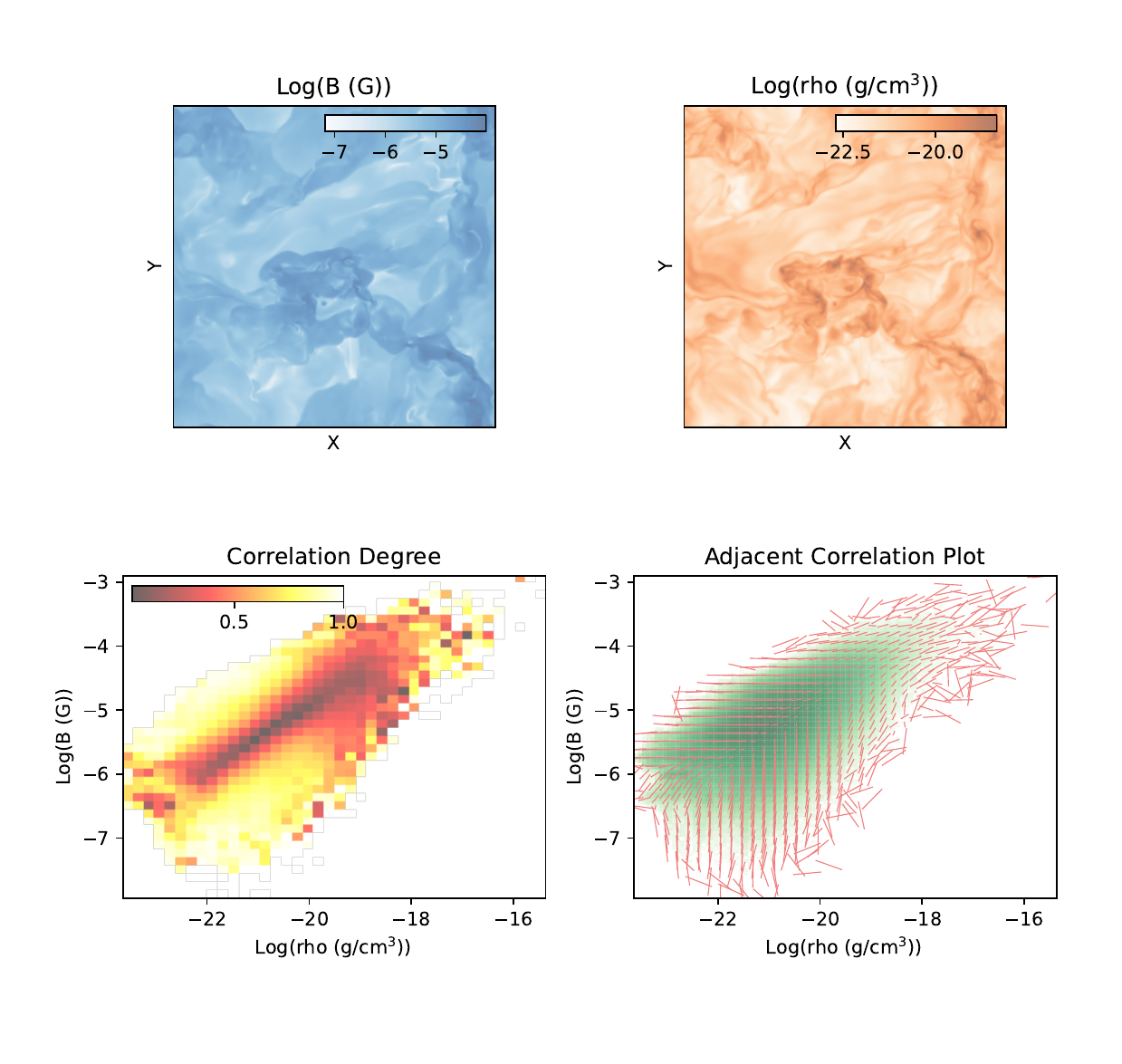}
\caption{{\bf Application of \emph{adjacent correlation analysis} to magnetized turbulence}. In the upper panels, we plot the density $\rho$ and magnetic field $B$ from the simulation. In the lower-left panel, we plot the correlation degree, and in the lower-right panel, we plot the density PDF with the local correlations represented using bars.  \label{fig:full:correlation} } 
\end{figure*}

 \section{Adjacency-Induced Correlations}

 


\emph{Adjacency-Induced Correlation} refers to the phenomenon where the relationship of values measured in adjacent locations (pixels/voxels) often contain correlations that are much stronger when compared to the global distribution. To illustrate this, we use a numerical simulation of magnetized supersonic turbulence \cite{2012ApJ...750...13C,2015ApJ...808...48B,2020ApJ...905...14B}, and plot the density against the magnetic field. The overall distribution in the magnetic field-density space is a cloud with a weak correlation. However, by choosing smaller boxes and plotting the values inside those boxes,  stronger correlations can be identified. We call this phenomenon the \emph{correlated variations of parameters from localized regions}, and provide means to describe this correlation.
This \emph{Adjacency-Induced Correlation} is a new type of regularity in data, which is hard to represent when producing the density plots.

\section{Adjacent Correlation Analysis}
\subsection{Correlation vectors}
 
The \emph{Adjacent Correlation Analysis} is a systematic approach to evaluate the correlated variations of parameters. 
We start with a few measurements,
 \begin{equation}
p_1(\vec{x}), p_2 (\vec{x}) \ldots p_i (\vec{x}) \;. 
\end{equation}
where $p_1, p_2, \ldotp p_i$ are different physical quantities, and  $\vec{x}$
represents the location. We select two quantities, $p_m$, and $p_n$, and plot
them against each other. This density distribution in the parameter space is
called the Probability Density Distribution (PDF).

To compute the local correlation vector, we identify all locations whose measurement values lie with the neighborhood of $(p_{m_i}, p_{n_i})$, evaluate the correlations at those individual locations, and sum all these correlations. 
 Toward each measurement,  the spatial gradients are
\begin{equation}
 \hat{\ {\vec{G_i}}} =  (G_{m,i}, G_{n,i}) =\left( \frac{\partial p_{m, i}}{\partial \vec{x}}, \frac{\partial p_{n, i}}{\partial \vec{x}} \right) \;,
\end{equation}
where $\frac{\partial p_{m, i}}{\partial \vec{x}}$ are directional derivatives.
$\vec{x}$ can be a set of directions.
 In e.g.3D, where the coordinates of $(x, y,
z)$, we choose $(\vec{e_x}, \vec{e_y}, \vec{e_z})$.


The correlated variation is the sum of the correlated variation vectors at different locations
\begin{equation}\label{eq:sum}
 \hat{\vec{C}} = \sum_i   \hat{\vec{G_i}}  w_i /  \sum_i w_i\;,
  \end{equation}
where $w_i$ (optional) is a weighting factor.
Since the correlation vector has a 2-fold symmetry (i.e. a vector rotated by 180$^\circ$ is identical to itself, which, in essence, is similar to the superposition of polarized light), the sum performed in Eq. \ref{eq:sum} must follow the role of the spin-2 vectors. The sum can be achieved using the linear components of Stokes parameters 
\cite{stokes1851composition} (see Methods \ref{sec:stokes}). For  each correlation vector, we have the following quantities:
\begin{itemize}
    \item Mean correlation degree $p$, ($0\leq p \leq 1$), and
    \item Correlation angle $\theta$.
\end{itemize}
These quantities can be overlaid on the PDF plot as vectors, forming the \emph{Correlated Variation Plot}. The procedure is illustrated in Fig. \ref{fig:illus}.  A version of the code is available at \url{}.
For a system containing $n$ variables, $p_1, p_2, \ldots, p_n$, the correlation we describe can be computed using a correlation matrix ${M}_{ij} = <(p_i - \bar{p_i} )(p_j - \bar{p_j})>$.

\subsection{Regularity measure and system classification}
We define the quantity $R$ to define the regularity contained in local correlations. The regularity $R$ is defined as 

\begin{equation}
 R  = \frac{\int \rho p {\rm d} v}{ \int \rho {\rm d} v} \;,
\end{equation}
where $\rho$ is the Probability Density, $p$ is the degree of coherence, and ${\rm d} v$ is the integration of the parameter space. This regularity measure describes the correlations that can be observed on the microscopic level in the parameter space.





\section{Applications}
We apply the adjacent correlation analysis to data from several systems.
The regularity $R$ of the systems we studied are presented in Table \ref{tab:regularity}, where the data we study all exhibit different degrees of regularity.  Extensive descriptions of the systems we study can be found in Methods \ref{sec:examples}.

\begin{table}
    \centering
    \begin{tabular}{cc}
    
 System & Regularity \\
        \hline
        \\
        \multicolumn{2}{c}{MHD simulations} \\
        \hline
 log($B$) vs. log($\rho$)& 0.47 \\
         $B_x$ vs. $B_ys$ & 0.21 \\
        \\
        \multicolumn{2}{c}{Turing Pattern} \\
        \hline
         $U$ vs. $V$ & 0.81 \\
         $V$ vs. $\nabla^2 V$ & 0.84 \\
        \\
        \multicolumn{2}{c}{Lorentz model} \\
        \hline
        $x$ vs. $y$ & 0.87 \\
        \\

        \multicolumn{2}{c}{Real-world data} \\
        \hline
 Climate data, Log(Precipitation) vs. temperature & 0.75 \\
 Molecular cloud, log(velocity) vs. log(density) &0.25 \\
    \end{tabular}
    \caption{{\bf Regularity measure of the systems studied.} The regularity measure $R$ is defined to describe the degree of coherence in the parameter space. }
    \label{tab:regularity}
\end{table}


\begin{figure*}
    \includegraphics[width=1 \textwidth]{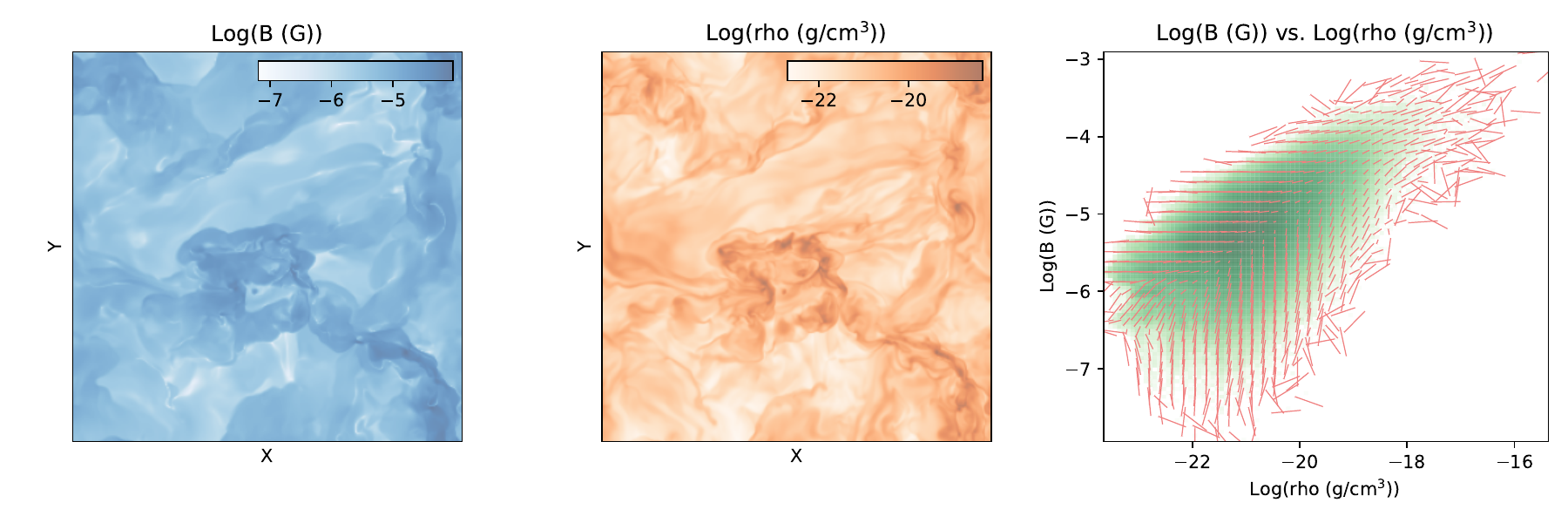 }
    \includegraphics[width=1 \textwidth]{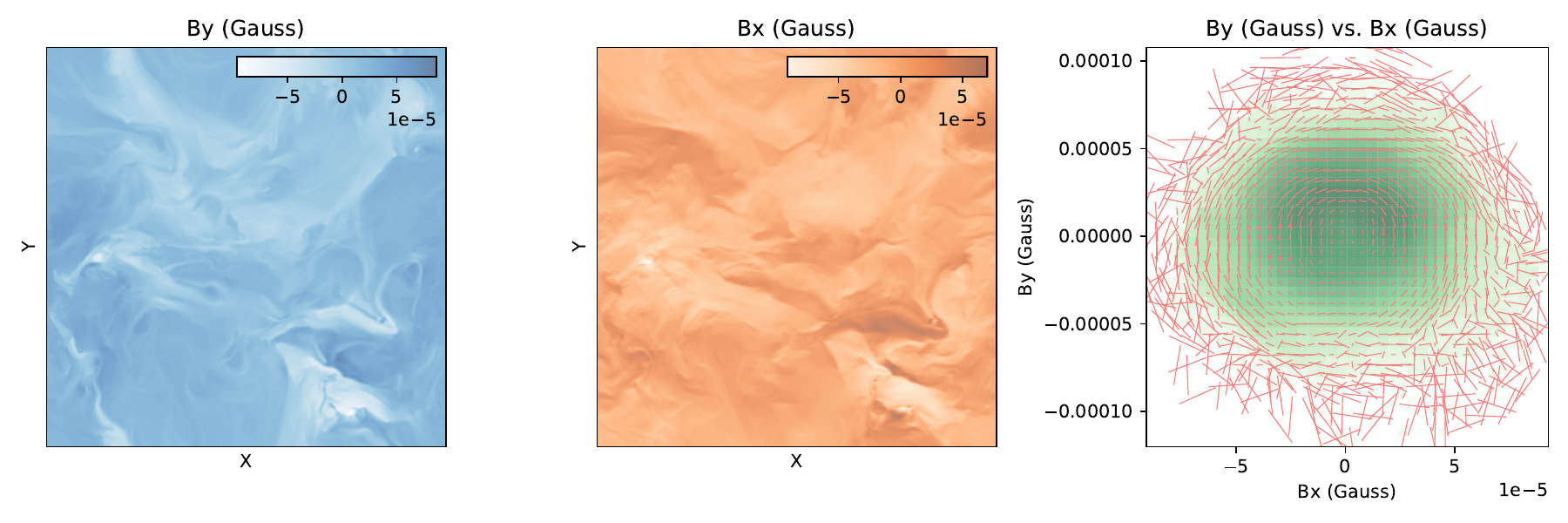}
    \caption{ {\bf Results from a simulation of MHD turbulence.} In the left and middle panels, we plot the quantities of interest, and in the right panels, we plot the results from the \emph{adjacent correlation analysis}. \label{fig:result:mhd}  }
\end{figure*}

\subsection{Simulations of compressive MHD turbulence}
We use numerical simulations of magnetized turbulence and plot the density against the magnetic field. Turbulence is a complex, multi-scale system that is notoriously difficult to analyze, and the most commonly used tools in previous turbulence studies include the correlation function and the Probability Density Distribution (PDF). Using a simulation of magnetized, supersonic turbulence, we demonstrate how the adjacent correlation analysis can reveal new regularities overlooked by previous approaches.

The application of the \emph{adjacent correlation plot} the widespread existence of Adjacency-Induced Correlations (Fig. \ref{fig:full:correlation}). Based on the system's behavior measured in such correlations, we can further segregate the data into different regimes of different properties. On the upper left side, we are in a regime of strong magnetization where the local variations are dominated by strong density variation upon a mostly uniform magnetic field. On the other side, we observe regions where variation in the magnetic field dominates. From the map of the correlation degree $p$, one can see that these two regimes have a high degree of correlation, separated by a region of low correlation.

The adjacent correlation plot is a flexible method that can be applied to explore the correlations between any two variables. In Fig. \ref{fig:result:mhd}, we study the correlation between the $x$-component of the magnetic field, $B_x$ against $B_y$, and discover a circular-shaped pattern of correlations, where, along each circle, we have $B_{xy} = \sqrt{B_x^2 + B_y^2} \approx \rm  constant$. A tentative explanation for this behavior is that this circular pattern results from the contributions of eddies with different energies.  

The regularity measure also offers some additional information on the interpretation of the plots: the ${\rm log}B-{\rm log} \rho$ plot has a regularity measure of 0.47, which is higher than the $B_x$-$B_y$ plot $(R= 0.21)$. The higher regularity in the  ${\rm log}B-{\rm log} \rho$ indicates a clear separation of the system into different regimes,  yet the lower $R$ in the $B_x$-$B_y$ plot is the result of  ${\rm log}B-{\rm log} \rho$ plot is understandable since the circular pattern probably results from the contributions of eddies of imperfect shapes.  



\begin{figure*}
    \includegraphics[width=1\textwidth ]{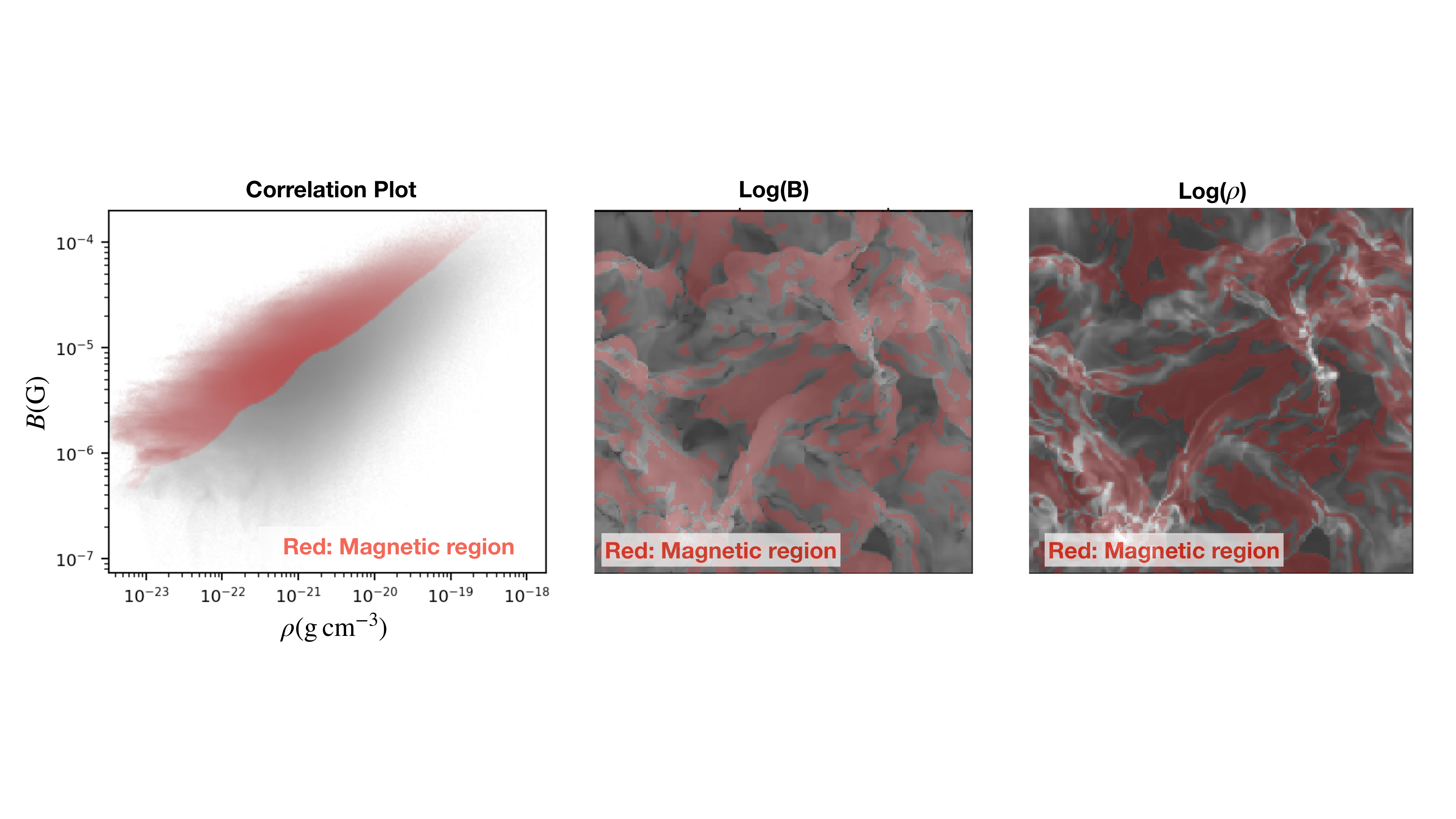}
\caption{{\bf Example of interactive data analysis.} We analyze the logarithm of the density and the magnetic field from a simulation of magnetized supersonic turbulence. The left panel shows the density against the magnetic field. Using software such as \texttt{Glue} \url{https://glueviz.org/}, one can select particular regions in the parameter space, and study the spatial distributions. Here, we have selected regions where the density variations dominate and found that spatially, they are distributed in large, coherent patches. \label{fig:interactive}}
\end{figure*}

\subsection{Interactive data exportation}
One practical use of the adjacent correlation analysis is to perform interactive data visualization. Although the method can effectively represent local corrections in the parameter space, the full map between spatial distribution and the parameter space can only be revealed through an interactive approach.

In Fig. \ref{fig:interactive}, we show an example of interactive data analysis using the software \texttt{Glue} \url{https://glueviz.org/} \cite{2015ASPC..495..101B,2017zndo...1237692R}. We analyze the logarithm of the density and the magnetic field from a simulation of magnetized supersonic turbulence. The parameter space can be divided into different regimes \cite{2024arXiv240902769L}: the magnetize regime, on the upper left, and the kinetic region, on the lower right (see Fig. \ref{fig:full:correlation}). 
We have selected the parameter range where the density variations dominate and found that they correspond to a few large, coherent patches. These are regions where the magnetic energy dominates over the kinetic energy, such that the local energy fluctuations from compressible turbulence are related to the $B \cdot \delta B$ term in the MHD equation \cite{2023A&A...672L...3S}. The fact that one can divide a domain into regimes based on how parameters correlate locally is similar to the \emph{data-driven balance model}\cite{2021NatCo..12.1016C}. The difference is our correlation-based approach, though not as rigorous, is more flexible and easily applicable.

\section{Interpretation of adjacent correlations}
The achievement of our adjacent correction analysis is the additional information it can extract in addition to the density PDF. The orientation of the local correlations often deviates significantly from the global correlations, thus carrying complementary information. Experts with established knowledge would start the intercalation of adjacent correlation plots with ease. Here, we discuss the conceptual value of the adjacent correction vectors. 





\begin{figure*}
    \includegraphics[width=1\textwidth ]{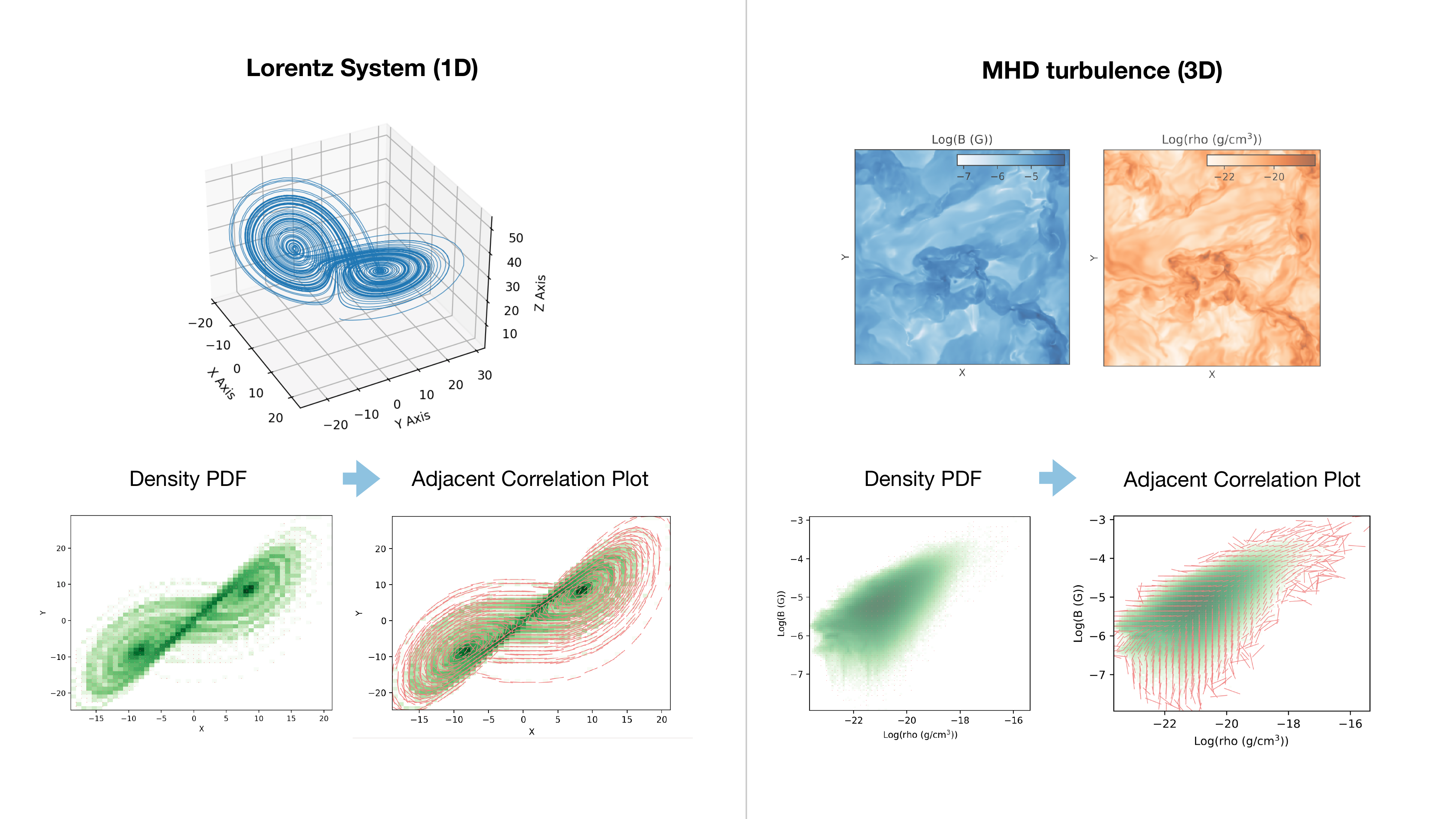}
\caption{{\bf Analogy between dynamical systems and patterns.} The left panel shows the evolution trajectory of the Lorentz system in the $x-y-z$ space.  When observed in the $x$-$y$ space, the vector field  (red bars) can be used to forecast the system's short-term behavior. The right panel shows the correlation vectors derived from the adjacent correlation analysis towards a simulation of magnetize, supersonic turbulence, where the vector field we derived provides some additional information, similar to the vector field in the Lorentz system. \label{fig:compare}}

\end{figure*}

\subsection{Spatial patterns as dynamical systems}

The correlation vectors we derive allow us to view patterns in nature using a language similar to that of the dynamical systems. 
The concept of phase space is crucial for understanding the evolution of dynamical systems, with the behavior of a dynamical system completely determined by a vector field in the phase space. In a significant fraction of the cases, the evolution of the system is restricted to some low-dimensional manifolds, where knowledge of the shape of these manifolds, as well as the topological structure of the vector field on top, can be used to predict the behavior of the system. One classical example is the Lorenz system  \cite{lorenz1963deterministic}, which is the earliest system with chaotic behavior being observed. The system is controlled by the following equations,

\begin{eqnarray}
 \frac{dx}{dt} &=& \sigma (y-x) \nonumber \\
 \frac{dy}{dt} &=& x(\rho - z) - y \nonumber \\
 \frac{dz}{dt} &=& xy - \beta \nonumber z \;,
\end{eqnarray}
where the chaotic behavior is characterized by hopping between the two lobes in the phase space. One crucial realization from the dynamical system research is that although it is nearly impossible to predict long-term evolution, short-time evolution can be predicted through its location in the phase space using the vector field. This can be achieved even in cases with only a limited number of observables, e.g. in the $x-y$ space (Fig. \ref{fig:compare}).

The vector field we derived using the \emph{adjacent correlation analysis} (Fig. \ref{fig:compare})  provides a similar description to patterns, where the correlation vector we derive occupies a similar location to the vector field in a dynamical system, with the difference that here the vector field has the 180$^{\circ }$ rotational symmetry. The adjacent correlation analysis can be used to estimate the short-range behavior of the system.

\subsection{Emergence of locally-conserved properties}
In a significant fraction of the cases, the emergence of locally correlated variations can be attributed to the emergence of locally conserved quantities after the system has reached 
 asymptotic states. 
The Buckinghum's pi theorem states that any physical systems can be written as 
\begin{equation}
 f(\Pi_1, \Pi_2, ...,\Pi_i) = 0 \;,
\end{equation}
and under special cases where some parameters are very small, we can drop them from the equations, such that the governing equation
\begin{equation}
 f(\Pi_1) = 0 \;,
\end{equation}
which means
\begin{equation}
    \Pi_i = q_1^{\alpha_1} \; q_2^{\alpha_2} , ...,q_i^{ \alpha_i} = C \;.
\end{equation}
In the logarithm space, we have
\begin{equation}
    \alpha_1 {\rm log}(q_1) +   \alpha_2 {\rm log}(q_2) + \ldots + \alpha_i {\rm log}(q_i)  = {\rm log}(C) \;,
\end{equation}

thus

\begin{equation}
    \alpha_1 {\rm d } ({\rm log}(q_1)) +   \alpha_2 {\rm d }({\rm log}(q_2)) + \ldots +  {\rm d } (\alpha_i {\rm log}(q_i))  = {\rm log}(C) \;,
\end{equation}
which should appear as a correlation had the appropriate parameters been chosen.


\begin{figure}
    \centering
    \includegraphics[width=1 \textwidth]{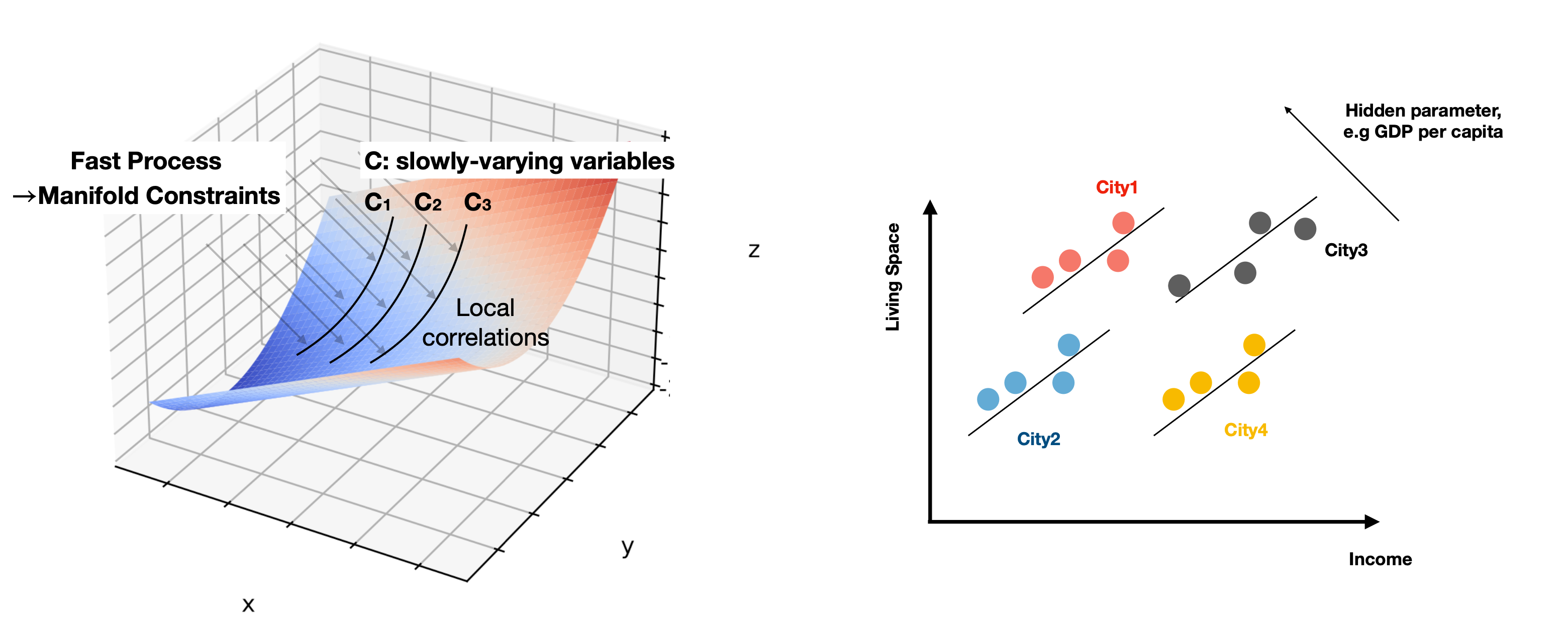}
    \caption{\label{fig:analogy}{\bf Left:}
     Manifold structure and Adjacent Correlation Analysis. {\bf Right:} llustration of the emergence of local correlations in a 3-parameter systems.}
\end{figure}


\subsection{Manifold interpretation} 
We briefly discuss how such a manifold view can improve our understanding of spatial patterns.
One important realization from the dynamical system is the phenomenon of the separation of scales \cite{doi:10.1137/151004896}, where one can write predictive equations in terms of a smaller set of state
variables. This situation typically arises in systems containing several disparate time scales;
if we are interested in studying the long-term dynamics, we may not want/need to resolve the fast time scales since they should already be saturated, nor do we need to resolve the slow times, as their effect should be minimal. 

In a manifold view, for a system controlled by a series of PDEs, a fast process will restrict the system to a manifold, where the local variations can be described by a (spin-2) vector field. The existence of some slow variables ($C$) might serve the role of separating different trajectories, which correspond to different spatially coherent regions. 
This view explains many phenomena observed in the adjacent correlation analysis since what we see in the adjacent correlation plot is a projected view of the attracting manifold. This analogy is illustrated in Fig. \ref{fig:analogy}. 

Consider the correlation between income and the size of the apartment, when measured in localized regions, families with larger incomes tend to live in larger apartments, and families with smaller incomes tend to live in smaller apartments. However, when we consider the whole country, the correlation between income and apartment size is weak. This is because the size of the apartment is not a direct function of the income but also depends on other hidden parameters, such as GDP per capita, and the hidden parameter must be slowly-varying (Fig. \ref{fig:analogy}). 

This simplified scenario explains the fact that when analyzing these correlations when evaluating the correlations between the two parameters locally, one can observe a strong correlation, but when the data is viewed globally, the correlation is weak, and this local correlation can be retrieved either using the adjacent correlation analysis or by setting the slowly-changing 3rd parameter (e.g. the GPD per capita) as constants. The lines the adjacent correlation plot reveals can be viewed as iso-surfaces on the latent manifold, with the hidden parameter constrained to constants.

\section{Discussions}
Discovering regularities in data is a central challenge in modern research. While data-driven and machine learning-based approaches have gained prominence, many analyses still rely on plotting data across parameter spaces to identify correlations, often overlooking spatial information.

We propose an effective method to leverage spatial information by analyzing correlations of parameter values in adjacent regions. Our observations indicate that quantities measured in local regions exhibit stronger correlations than those measured across entire datasets. To capture these local correlations, we developed a novel approach termed adjacent correlation analysis, with results visualized as adjacent correlation plots.

Adjacent correlation analysis offers a powerful tool for data exploration, revealing insights beyond those provided by probability density distribution plots. While density plots provide distributions in the $p_1$-$p_2$ plane, adjacent correlation analysis quantifies the correlation degree $p$ and angle $\theta$, revealing structured patterns in correlation vector maps. For instance, circular patterns emerge in $B_x$-$B_y$ plots, and distinct vector orientations appear in ${\rm log}(B)$-${\rm log}(\rho)$ plots. These structures facilitate the discovery of data-driven laws, such as locally conserved quantities. Additionally, adjacent correlation plots enable quantitative comparisons of spatial patterns by aligning their correlation vectors, serving as a robust foundation for interactive data analysis.

Adjacent correlation analysis establishes a conceptual link between the phase space of a dynamical system and the phase space of patterns, replacing the vector field of the dynamical system with spin-2 correlation vectors. This analogy enables the application of dynamical systems concepts, such as attracting manifolds and their associated vector fields, to interpret the results of adjacent correlation analysis. In the long term, this approach should allow researchers to adapt tools from dynamical systems theory to study pattern formation.

Several extensions to the method are under development. For systems exhibiting spatial anisotropy, correlations can be analyzed separately along different directions rather than summed, enhancing the method’s flexibility. Given that many physical systems are governed by partial differential equations, incorporating temporal derivatives and comparing them with spatial results may uncover new data regularities. The current implementation is optimized for two-dimensional parameter spaces, facilitating interactive data analysis. However, extending the method to higher dimensions is essential. Additionally, generalization to irregularly sampled data is in progress and will be made available in future iterations.



\section*{Acknowledgements}
GXL acknowledges support from,
NSFC grant No. 12273032 and 12033005. GXL would like to thank his like-minded colleagues without whom this project would not be possible (dynamical systems: Prof. Xun Shi, patterns in simulation data: Dr. Mengke Zhao, information, computation, and free exploration: Prof. Torsten Ensselin, and commonality among different astrophysical systems: Prof. Douglas N. C. Lin). GXL also thank 
 Prof. Andrew Laszlo for sharing his view on scale separation and recommendation of the Transtrum et al. paper that the authors would not find otherwise.  Qiqi Jiang is acknowledged for recommending the Jain \& Haller paper.

\section*{Author contributions statement}
GXL designed the project, performed the analysis and wrote the paper. 

\section*{Code Availability}
The code of the method can be found at \url{https://github.com/gxli/Adjacent-Correlation-Analysis}.

\bibliography{paper}

\appendix

\section{Sum of spin-2 vectors using the Stokes parameter }\label{sec:stokes}
Our measurements are ($E_{x,i}, E_{y,i}$). To perform the sum, we first convert them into the \emph{Stokes parameter} 
\begin{eqnarray}
    I_i &=& E_{x,i}^2 + E_{y,i}^2 \\ \nonumber
    Q_i &=& E_{x,i}^2 - E_{y,i}^2\\ \nonumber
    U_i &=& 2 E_{x,i}\, E_{y,i}\\ \nonumber
    V_i &=& 0
\end{eqnarray}
and the sums are
\begin{equation}
    I = \sum_i w_i I_i, \quad Q = \sum_i w_i Q_i, \quad U = \sum_i w_i U_i, \quad V = \sum_i w_i V_i
\end{equation}
The degree of correlation is
\begin{equation}
    p = \left( \left( Q/I\right)^2 + \left(U/I\right)  \right)^{1/2}
\end{equation}
and the position angle is 
\begin{equation}
    \theta = \frac{1}{2} \arctan \left( \frac{U}{Q} \right)
\end{equation}

\section{Other Examples}\label{sec:examples}

\begin{figure*}
    \includegraphics[width=1 \textwidth]{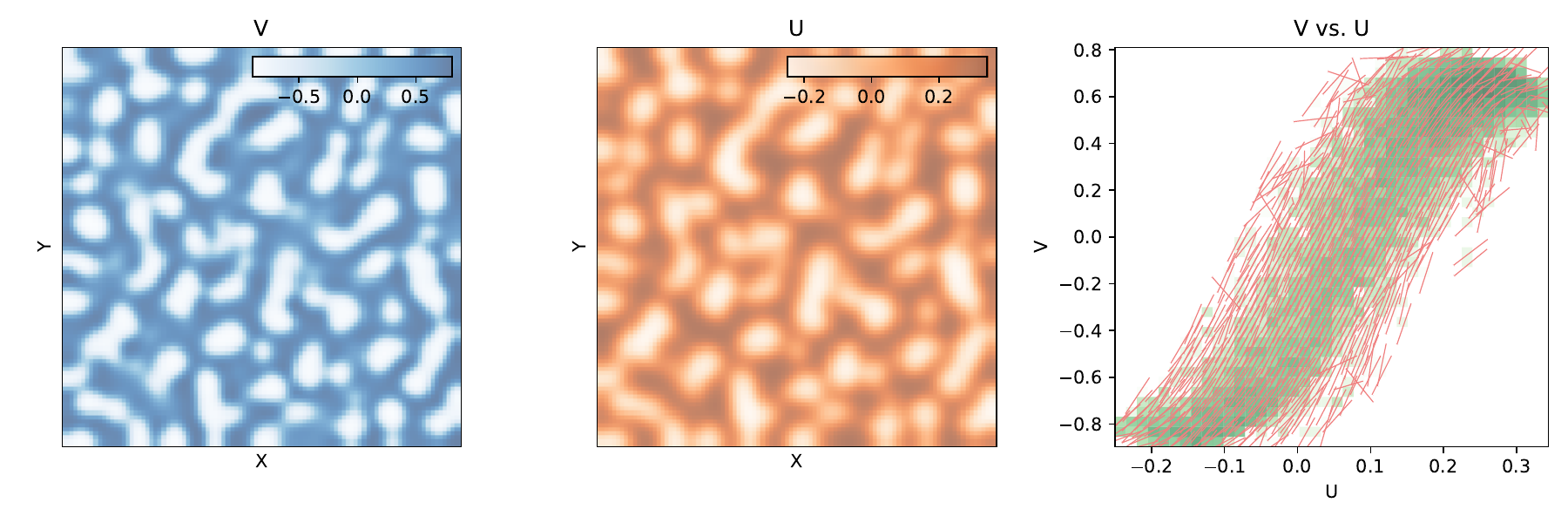}
    \includegraphics[width=1 \textwidth]{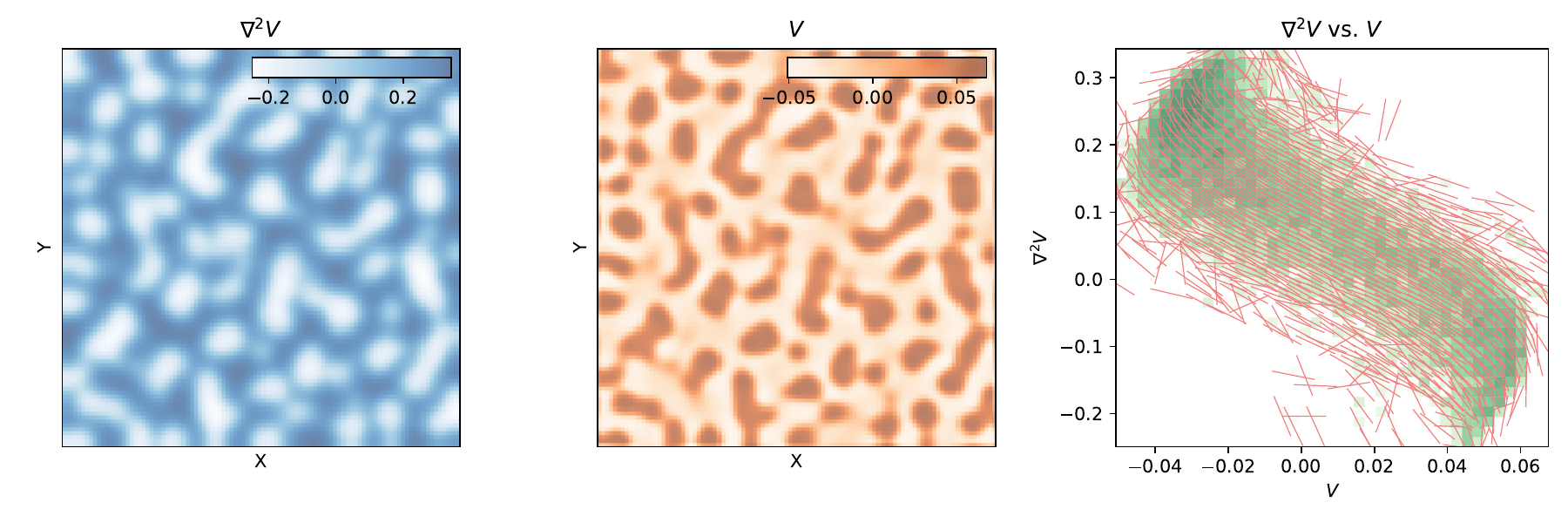}

    \caption{\label{fig:result:turing} {\bf Results from a simulation of Turing pattern.} In the left and middle panels, we plot the quantities of interest, and in the right panels, we plot the results from the \emph{adjacent correlation analysis. }}
\end{figure*} 

\begin{figure*}
    \includegraphics[width=1\textwidth ]{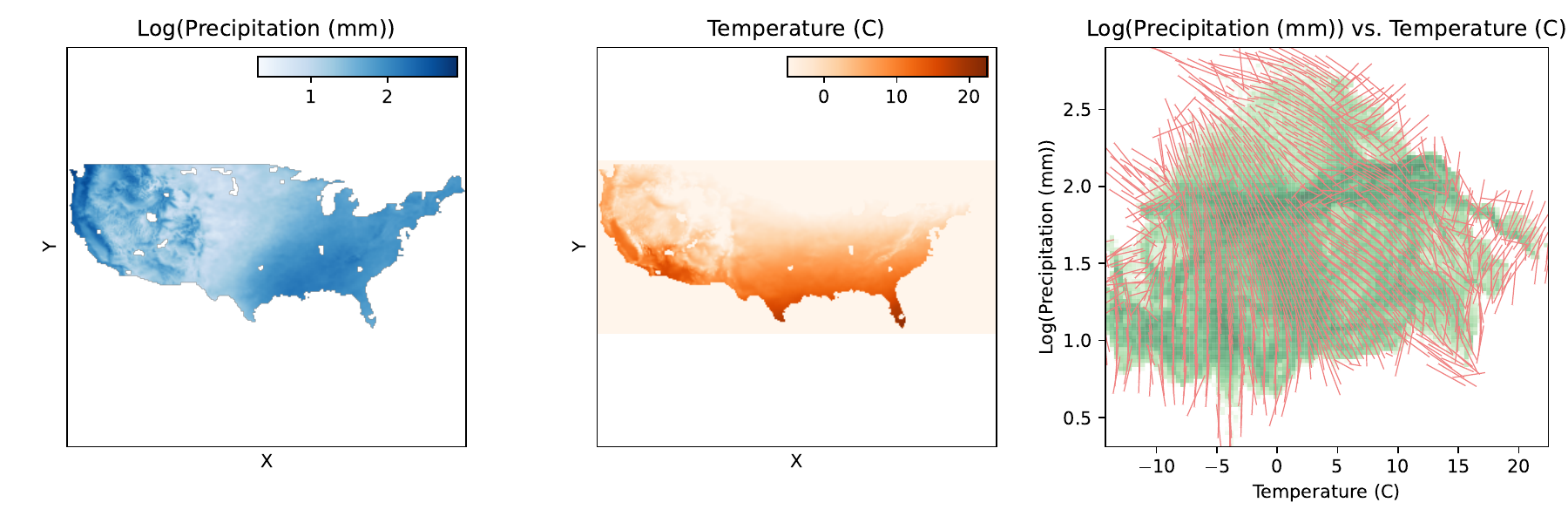}
    \includegraphics[width=1\textwidth ]{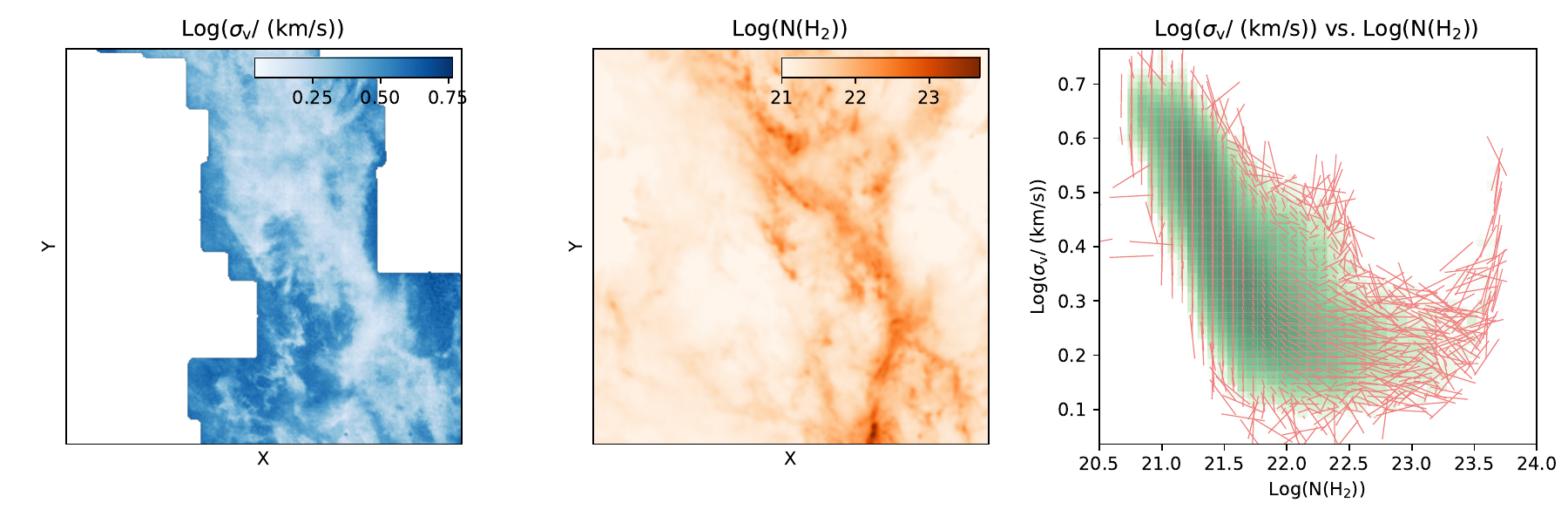}
\caption{{\bf Application of the adjacent correlation analysis to real-world data.} In the upper panels, we plot the result of the adjacent correlation analysis applied to temperature and precipitation data obtained from NOAA, and in the lower panels we plot results from observations towards the Orion molecular cloud, where we study the relationship between the velocity dispersion $\sigma_{\rm v}$ and the H$_2$ surface density $N_{\rm H_2}$. \label{fig:result:realworld }}
\end{figure*}

\subsection{Turing pattern from reaction-diffusion systems}
The Turing pattern\cite{1952RSPTB.237...37T} is a concept that describes how patterns such as stripes and spots can arise naturally and autonomously from a homogeneous, uniform state. It is one of the earliest systems where patterns emerge from a set of well-defined differential equations. 
We simulate the formation of Turing patterns using
\begin{eqnarray} \nonumber
 \frac{\partial u}{\partial t} &=& a \nabla^2 u + u  - u^3 -v + k \nonumber \\
   \tau \frac{\partial v}{\partial t} &=& b  \nabla^2 v + u - v\;, \nonumber 
\end{eqnarray}
where $a=2.8 \times 10^4, b=5\times 10^{-3}, \tau=0.1$ and $k=-0.005$, and study the pattern at the system has reached a quasi-stationary stage. The results are plotted in Fig.  \ref{fig:result:turing}. In both the $U$-$V$ and $V$-$\nabla^2 V$ parameter space, we find that the application of the \emph{adjacent correlation plot}  do reveal additional correlations, which reflects the regularity of the system where $R=0.81$ and $R=0.84$. The shape and correlations in these distributions offer a new way to quantify and compare these patterns.

\subsection{Weather data: Precipitation vs. temperature }
To demonstrate the effectiveness of the adjacent correlation analysis in real-world data, we apply the method to meteorological data. We obtain the monthly-averaged temperature and precipitation data from National Oceanic and Atmospheric Administration (NOAA) \footnote{\url{https://www.ncei.noaa.gov/cdo-web/}}, and apply the \emph{adjacent correlation analysis} measured towards North America. The data we use are taken from January, measured over the period of 2006 to 2020. The results are plotted in Fig. \ref{fig:result:realworld }. The data exhibit strong correlations at different locations in the parameter space. This example illustrates the widespread existence of correlations in data, where we find $R=0.75$. In most of the areas, the local increase of the temperature is related to a decrease in the precipitation. We also observe that in very cold areas (lower left corner of the plot), the precipitation can fluctuate significantly without the temperature changing much. Based on the adjacent correlation analysis, we can further segregate the data into different regimes of different properties through some interactive explorations.

\subsection{Molecular cloud data}
As another example of real-world data, we apply the adjacent correlation analysis to molecular cloud data. We use the data from the Orion molecular cloud taken by CARMA-NRO Orion Survey \cite{2018ApJS..236...25K}, and plot the velocity against the density (Fig. \ref{fig:result:realworld }). The data exhibits a regularity measure of $R=0.25$, which is lower than the other examples. The lower value of the regularity measure is understandable since in these real-world systems, the observations are limited and imperfect. Thus, one does expect little regularity measure to fluctuate between systems, as the parameters one analyzes are not all expected to be correlated. Nevertheless, one can observe a clear positive correlation between the surface density and the velocity dispersion at the high-surface density end, which can be the signature of the gravitational collapse (where the increase of mass, as seen from the increase of the surface density, leads to collapse, as reflected in the increase of the velocity dispersion).

\end{document}